\documentclass[letterpaper, 12pt]{article}
\usepackage[utf8]{inputenc}
\usepackage[english]{babel}
\usepackage{amsmath}
\usepackage{amssymb}
\usepackage{amsfonts}
\usepackage{color}
\usepackage{graphicx}
\usepackage{hyperref}

\newtheorem{theorem}{Theorem}[section]

\title{Modelling epidemics: a perspective on mathematical models and their use in the present SARS-CoV-2 epidemic}
\author{Jorge X. Velasco-Hernandez\\
Instituto de Matemáticas Unidad Juriquilla\\
Universidad Nacional Autónoma de México, Quer\'etaro CP 76230}
\date{March 2021}

\begin{document}

\maketitle
\begin{abstract}
In this work we look at several mathematical models that have been constructed during the present pandemic to address different issues of importance to public health policies about epidemic scenarios and thier causes. We start by briefly reviewing the most basic properties of the classic Kermack-McKendrick models and then proceed to look at some generalizations and their applications to contact structures, cocirculation of viral infections, growth patterns of epidemic curves, characterization of probability distributions and passage times necessary to parametrize the compartments that are added to the basic Kermack-McKendrick model. All of these examples revolve around the idea that a system of differential equations constructed for an specific epidemiological problem, has as central and main theoretical and conceptual support the epidemiological, medical and biological context that motivates its construction and analysis. Fitting differential models to data without a sound perspective on the biological feasibility of the mathematical model will provide an statistical significant fit but will also fall short of providing useful information for the management, mitigation and eventual control of the epidemic of SARS-CoV-2.
\end{abstract}
\section{Introduction}
The epidemic of SARS-CoV-2 that started in late December 2019 in Wuhan, China, and became a pandemic by late January 2020, is still with us. Mathematical models have been used since the very early days to forecast and inform public health strategies and decisions, geared to control or, at least, mitigate the impact of the epidemic. Epidemic models became a notoriety both in the academic, public health and popular realms. 
The epidemics of infectious diseases are social phenomena whose dynamcs is strongly influenced by the inherent heterogeneity of human populations regarding age, gender, genetic profiles, physiology, preferences, education, socioeconomic level, religion, customs, etc. All of these factors induce high variability and uncertainity on the disease transmission process. To confront the epidemic, a large number of modellers used variants of the family of so called Kermack-McKendrick models.The Kermack-McKendrick models were first published in a series of papers in early decades of the XX century. These models are compartamental in nature and follow the fate of the individuals in the population where an infectious disease is present. One of the equations follows the prevalence of the disease which, in the most classical SIR model, describes a bell-shaped curve with a unique maximum and where both initial growth and final decline are exponential in nature. This model originally was used to describe the bubonic plague epidemic \cite{Kermack1927}. 
Anderson G. McKendrick was a scottish medical practitiones born by the end of the XIX century. He worked in Sudan and India where he thought himself mathematics \cite{Bacaer2011}. As a result form this experience he published a first version of his now famous model  \cite{Kermack1991a} which later in his life was further developed with the help of W.O. Kermack in a series of landmark papers. In 1927 William O. Kermack and McKendrick started a series of papers entitled  Contributions to the Mathematical Theory of Epidemics that together constitute the fundamental background of mathematical epidemiology \cite{Kermack1927,Kermack1991,Kermack1991b,Kermack1933,Kermack1937}.
Infectious disease are transmitted through contacts between susceptible and infectious individuals. The models that we will comment on in the following sections, assume that all individuals are exacltly equal and can be unequivocally classified into distinct classes that correspond to their physiological disease status: susceptibles, exposed, infectious, recovered, quarantined, etc.  
During this long pandemic of SARS-CoV-2 very many papers have been published where many more compartments are added to the basic transmission model (hospitalized, asymptomatically infeted, presymptomatically infected, isolated individuals and so on). This exacerbated increase in the number of compartments gives the illusion of realism but at a price: the number of parameters, about two per new compartment, also increases bringing the necessity of either estimating all parameters or assigning values to some of them for the purpose of inference, simulation, forecast and prediction. Lack of identifiability is, however, the result and also the problem of overfitting arises. Many of the parameter values are taken from the medical and epidemiological literature and assumed to be true. This form of `parameter reduction´ does not necessarily diminishes the sensitivity of the model solutions to variations in these parameter values. Mathematical models are tools to study natural phenomena. Models can certainly be used to predict but they can also be used to explain and understand the mechanisms and causes of a given natural phenomenon. Kermack-McKendric models are excellent strategical tools of this type. Their objective is then not so much to forecast incidence, deads and hosptalized individuals,for example, but to generate scenarios and explanatory hypothesis that may be of help for those enaged in model forcasting. Before concluding this section, we point out that, in this pandemic, plenty of data has been made available to the general public since the initial stages. An epidemic is a biological and social process that first has to be understood in those terms. The analysis of this information for each national, regional, local epidemic is essential to understand and correctly design mathematical models with the technical power and biological feasibility and consistency that would make them useful tools to design and execute adequate public policies.  Fitting epidemic curves to models is only a part of it.
\section{Basic results of the Kermack-McKendrick model}
Mathematical models have been deployed during the present sanitary emergency in a variety of ways but mainly to forecast its course. The basic mathematical model that has been used for such an aim is the classical Kermack-McKendrick SIR mathematical model. The SIR model is a compartamental system of differential equations that models disease transmission of communicable diseases by subdividing the population into three classes: the susceptible class that contains individuals that are not immune to the disease and are at risk of contagion; infected individuals that can also transmit the disease, that is, they are also infectious, and the class or recovered or death individuals. Recovered individuals are immune to the disease.
The models stands as follows
\begin{align}\label{eq:sir}
    \dot{S}&=-\beta S I,\\\nonumber
    \dot{I}&=\beta SI -\gamma I\\
    \dot{R}&=\gamma R,\nonumber
\end{align}
with initial conditions $S(0)=S_0$, $I(0)=I_0$ and $R(0)=R_0$ and where $\beta=\phi c $ is  the effective contact rate with $\phi$ the  prbability of infection per contact, and $c$ the number of contacts per capita per unit time of a typical individual in the population. Note that all individuals are exactly equal with respect to the disease. We will elaborate on this aspect of the model later in this contribution; $\gamma$ is the recovery rate. Note that in the absence of new infections
$I(t)/I_0=\exp(\gamma t)$, where $I(0)$ is the initial number of infectious cases, and therefore
$$\frac{1}{\gamma}=\int^{\infty}_0\gamma e^{\gamma t}dt$$
is the mean residence time in the infectious compartment. Summing the equations in \ref{eq:sir} we see that $S(t)+I(t)+R(t)=N$ constant for all $t$. Take $N=1$. Using the first and the second equation we can see that the solution curves can be found by solving
$$\frac{dI}{dS}=-1+\frac{\gamma}{\beta S},$$
and thus $I=1-S+\gamma \log[S/S_0)/\beta.$ We have the following Theorem \cite{Hethcote1976}:

\begin{theorem}
Let $S(t),I(t)$ be solutions of (\ref{eq:sir}). Define $R_0=\beta S_0/\gamma$. If $R_0\leq 1$, then $I(t)$ decreases to zero as $t\to\infty$; if $R_0>1$, then $I(t)$ first increases up to a maximum value equal to $1-\gamma/\beta-\gamma\log(R_0)/\beta$ and then decreases to zero as $t\to\infty$. The susceptible fraction $S(t)$ is a decreasing function, and the limiting values $S(\infty)$ is the unique root in $(0,\gamma/\beta)$ of the equation
$$1-S(\infty)+\log(S(\infty)/S_0)\frac{\gamma}{\beta}=0.$$
\end{theorem}

The SIR equations (\ref{eq:sir}) describe a single epidemic outbreak where demographic processes (births, deaths, immigration and emigration) do not play a role and where immunity is complete. An implicit assumption is that the behavior of the individuals that make the population is always constant and homogeneous. This means that the effectve contact rate is the same for all, that there are no groups of people with different contact rates and probability of infection. This is obviously an oversimplification but also something that a careful reading of the model equations leaves clear. 
When vital dynamics is introduced, the SIR model becomes
\begin{align}\label{eq:sirvd}
    \dot{S}&=\mu -\beta S I-\mu S,\\\nonumber
    \dot{I}&=\beta SI -(\gamma+\mu) I\\
    \dot{R}&=\gamma R-\mu R,\nonumber
\end{align}
where $\mu$ is the birth and death rate (assumed here equal). Since $R(t)=1-S(t)-I(t)$, note that this system has a biologically feasible interpretation in the triangle $\Delta$ bounded by the $S$ and $I$ axes and the line $S+I=1$. Given the initial conditions $S(0)>0$, $I(0)>$, $R(0)\geq 0$ we can establish the following Theorem \cite{Hethcote1976}
\begin{theorem}
Let $S(t),I(t)$ be solutions of (\ref{eq:sirvd}). Define $R_0=\beta S_0/(\mu+\gamma)$. If $R_0>1$ then there exists a globally stable endemic equilibrium point in $\Delta -\{(S,0): 0\leq S\leq 1\}$ given by $E_1=(1/R_0, \mu (R_0 -1)/\beta$. If, on the contrary $R_0\leq 1$ the disease-free equilibrium is globally asymptotically stable in $\Delta$.
\end{theorem}
To appreciate the effect of temporal immunity on disease dynamics let us suppose that immunity conferred by the disease is short lived. We must point out that given the absence of vital dynamics, this model is only valid for short time scales (as the original SIR model (\ref{eq:sir}) is). However, it serves our purpose of showing the characteristic oscillatory dynamics imposed by the reintroduction of recovered individuals into the susceptible class. The model is the following: 
\begin{align}\label{eq:sirti}
    \dot{S}&=-\beta S I+\omega R,\\\nonumber
    \dot{I}&=\beta SI -\gamma I\\
    \dot{R}&=\gamma I-\omega R,\nonumber
\end{align}

\begin{figure}
    \begin{tabular}{cc}
         A&B\\\includegraphics[width=6.5cm]{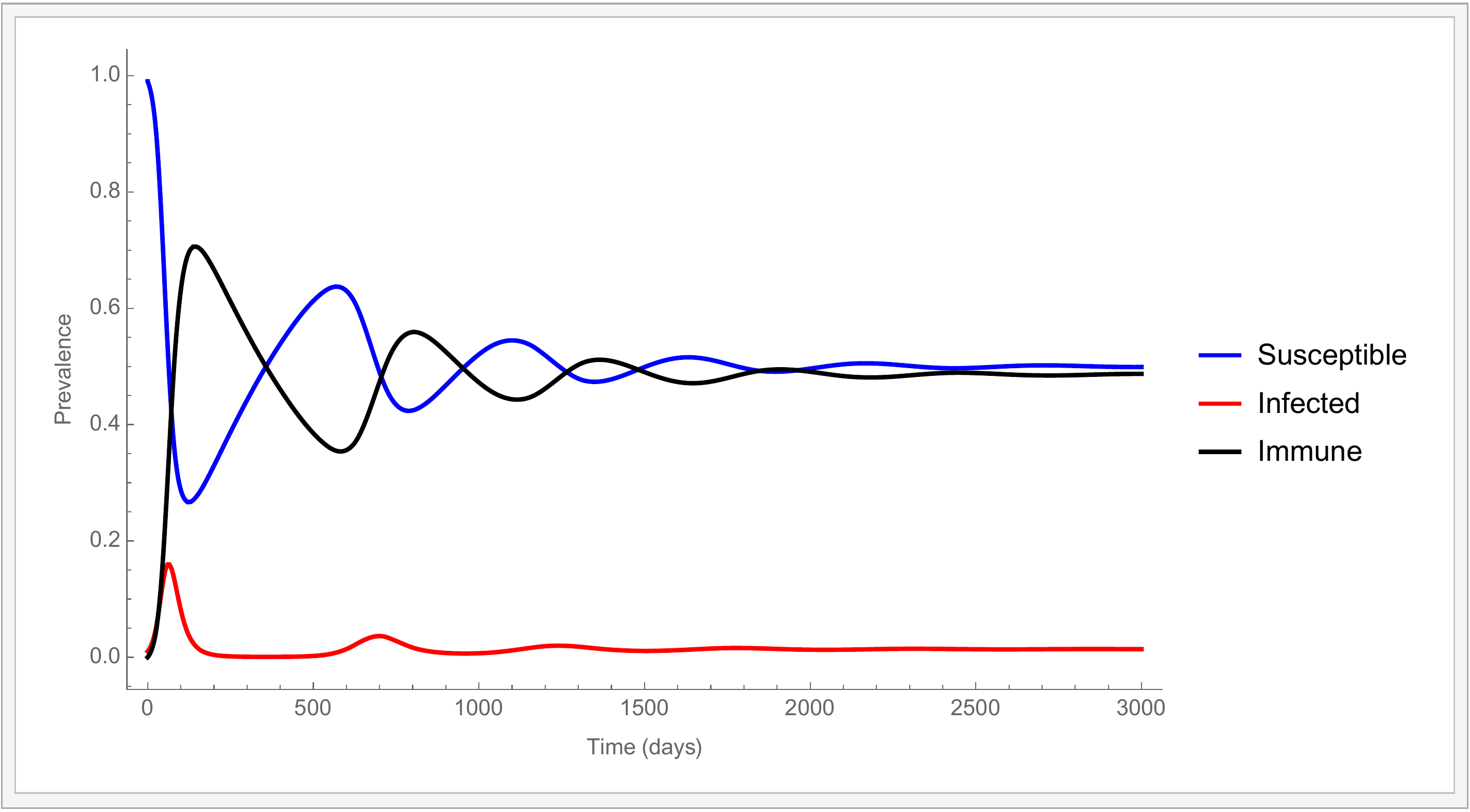}& \includegraphics[width=6.5cm]{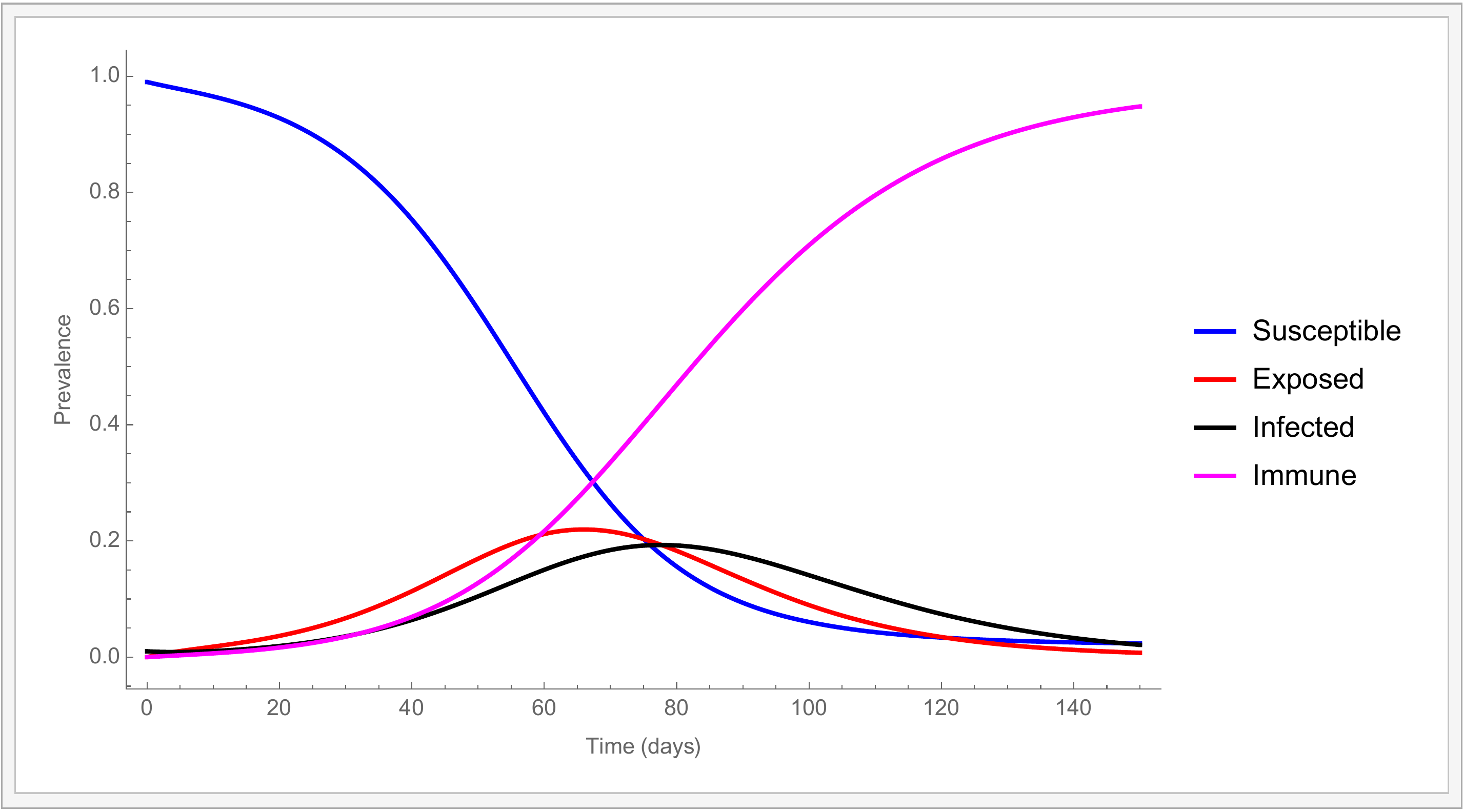}  
    \end{tabular}
    \caption{A) Dynamics of model (\ref{eq:sirti}), B)Dynamics of model (\ref{eq:seir}). }
  \label{fig:sirs}
\end{figure}

As mentioned above,  the dynamics shares characteristics with model (\ref{eq:sir}) in that we still have that if $R_0=\beta S_0/\gamma> 1$ the number of infected individuals will increase, reach a maximum when $S=\gamma/\beta$ and then decrease. However, notice that $S$ is no longer a nonincreasing function of time. The fact that immunity last only about $1/\omega$ units of time indicates that the susceptible pool is replenished as more people that where infected recover and then lose their immunity. This imposes a transient dynamic that consists of damped oscillations until de disease vanishes in the population (see Figure \ref{fig:sirs}).

To finish this preliminary introduction we present another Kermack-McKendrick model that has found multiple applications in the present pandemic. This is the  SEIR model given by:
\begin{align}\label{eq:seir}
    \dot{S}&=-\beta S I,\\\nonumber
    \dot{E}&=\beta SI -\eta I\\
    \dot{I}&=\eta I-\gamma E,\\\nonumber
    \dot{R}&=\gamma E,\nonumber
\end{align}
Here the only new parameter is $\eta$, the incubation rate; as before $1/\eta$ is the mean residence time that a person spends incubating the virus before becoming infectious. $E$ is a compartment that contains the exposed individuals defined as those that have been infected but are not yet infectious. It has similar properties to the SIR model but now presents two peaks: one corresponds to the prevalence of exposed and other to the prevalence of infectious individuals.  Note that the typical form of an epidemic outbreak, common to all three models, is a bell-shaped epidemic curve, that initially grows exponentially. For the SIR model (eq. \ref{eq:sir}) this can be easily illustrated. At the beginning of the epidemic we can assume $S$ is approximately constant and equal to $S_0$, thus, since $R_0=\beta S_0/\gamma$,
$$I(t)=I(0)e^{\gamma(R_0-1)t},$$
which allows us to define the population growth rate as $r=\gamma(R_0-1)$ which then implies that
$$R_0=1+\frac{r}{\gamma}\approx 1+\frac{T_R}{T_D},$$
where $T_R$ is the recovery time and $T_D$ is the doubling time of the epidemic, roughly equal to $1/r$. A similar expression can be found for the reproduction number of the SEIR model (eq.\ref{eq:seir}). Linearizing the system around the disease-free equilibrium and finding the dominant eigenvalue we obtain that
$$R_0=(1+\frac{r}{\gamma})(1+\frac{r}{\eta}).$$
Recall that a basic condition of the above two formulas strongly relies on the assumption that the probability distribution of the incubation and recovery periods is exponential
$$p_I(u)=e^{-\eta u},\qquad p_R(u)=e^{-\gamma u}.$$
This is only an approximation. In reality, the distributions of these times are Gamma distributions, not exponential, a fact that is important for the correct estimation of the basic reproductive number and the initial epidemic growth rate.

\section{More general models}
\begin{figure}
    \begin{tabular}{cc}
         A&B\\\includegraphics[width=6cm]{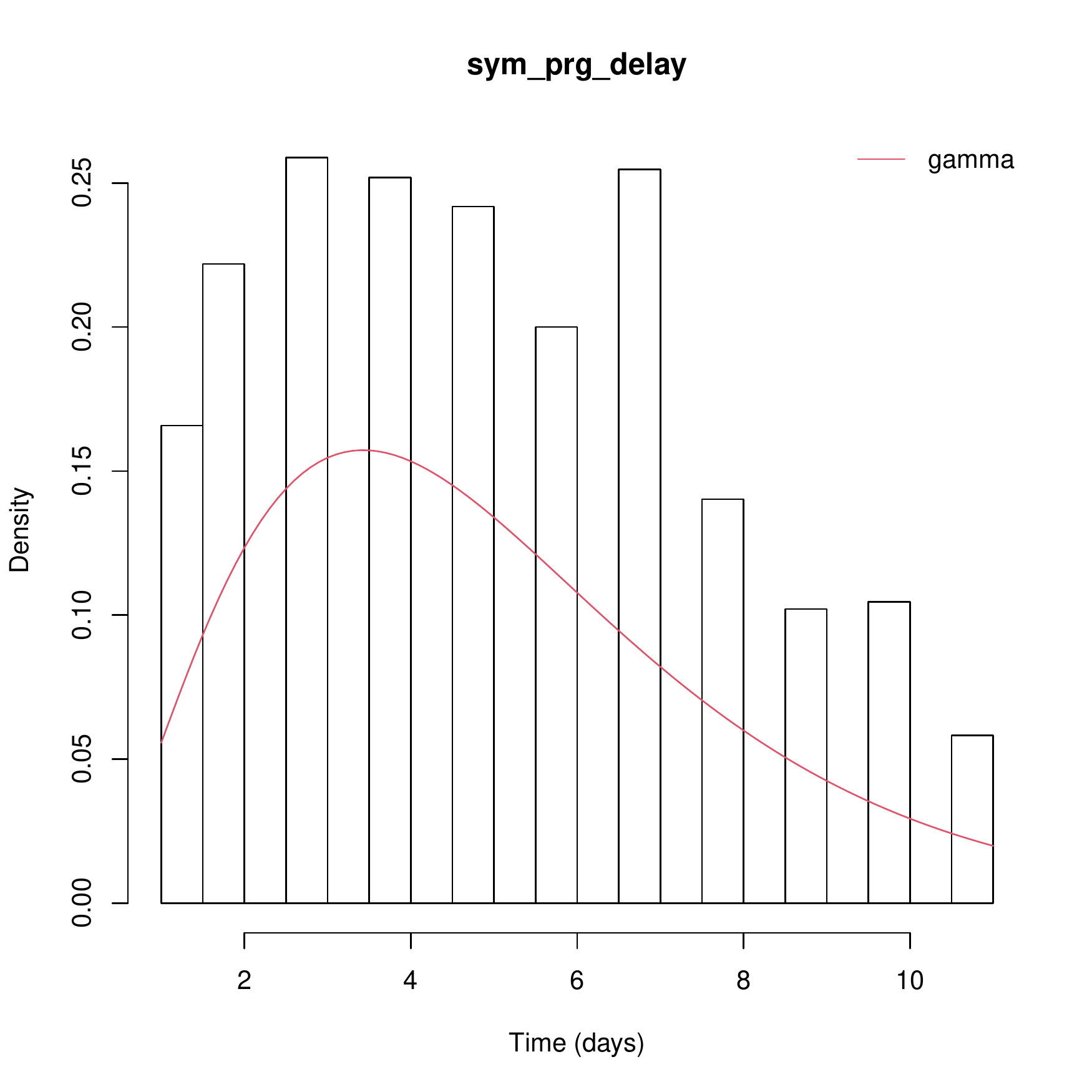}& \includegraphics[width=6cm]{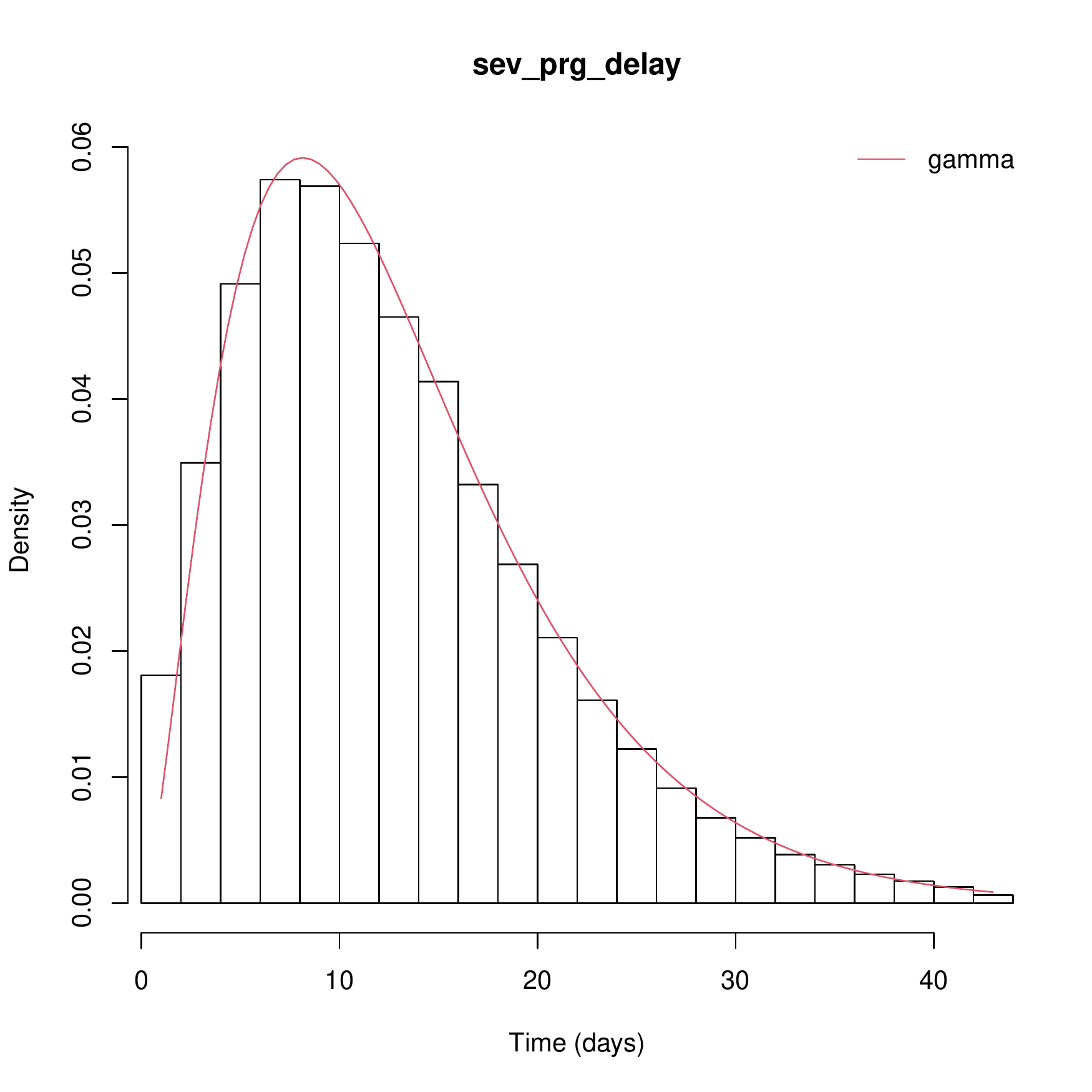}  \\
         C&D\\
         \includegraphics[width=6cm]{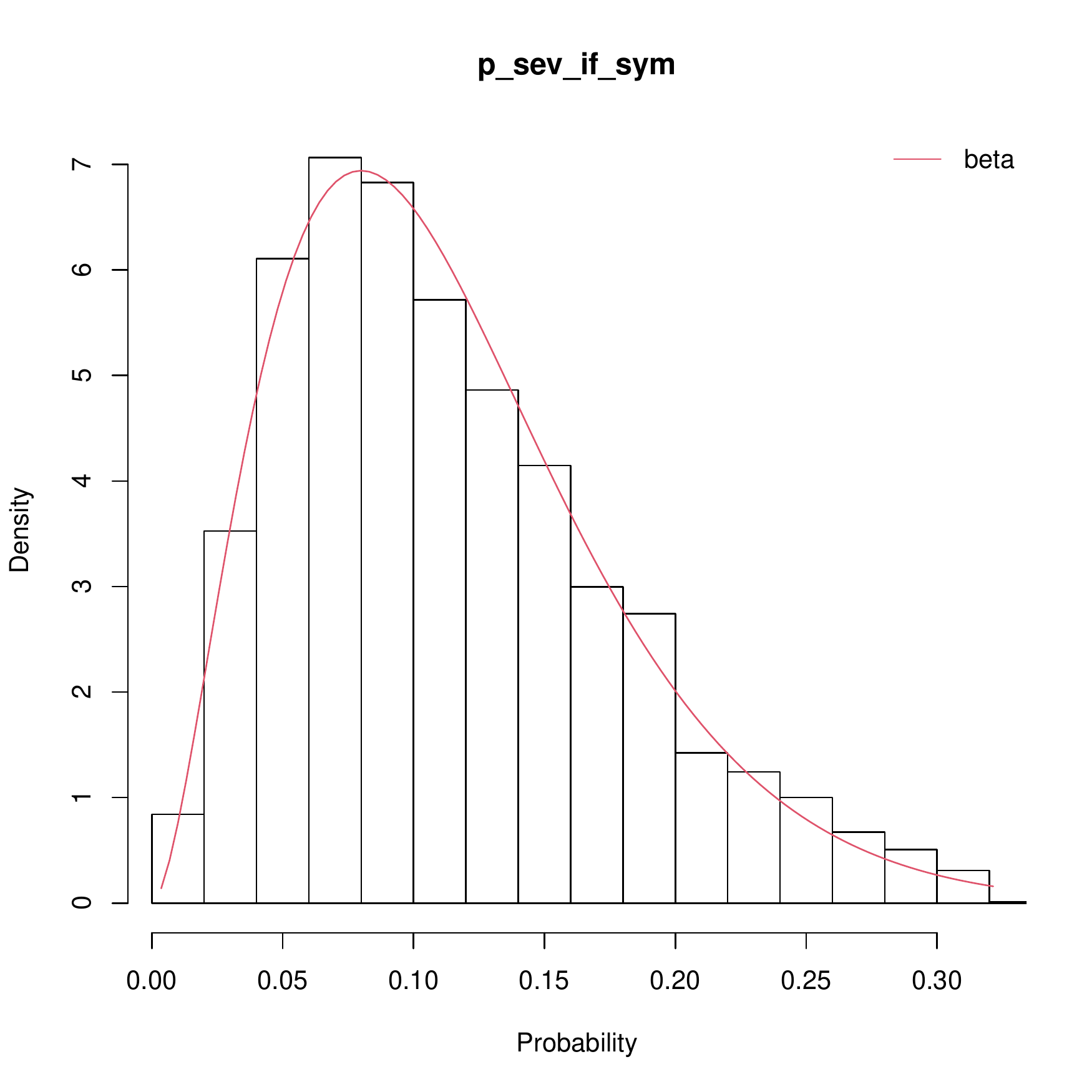} & \includegraphics[width=6cm]{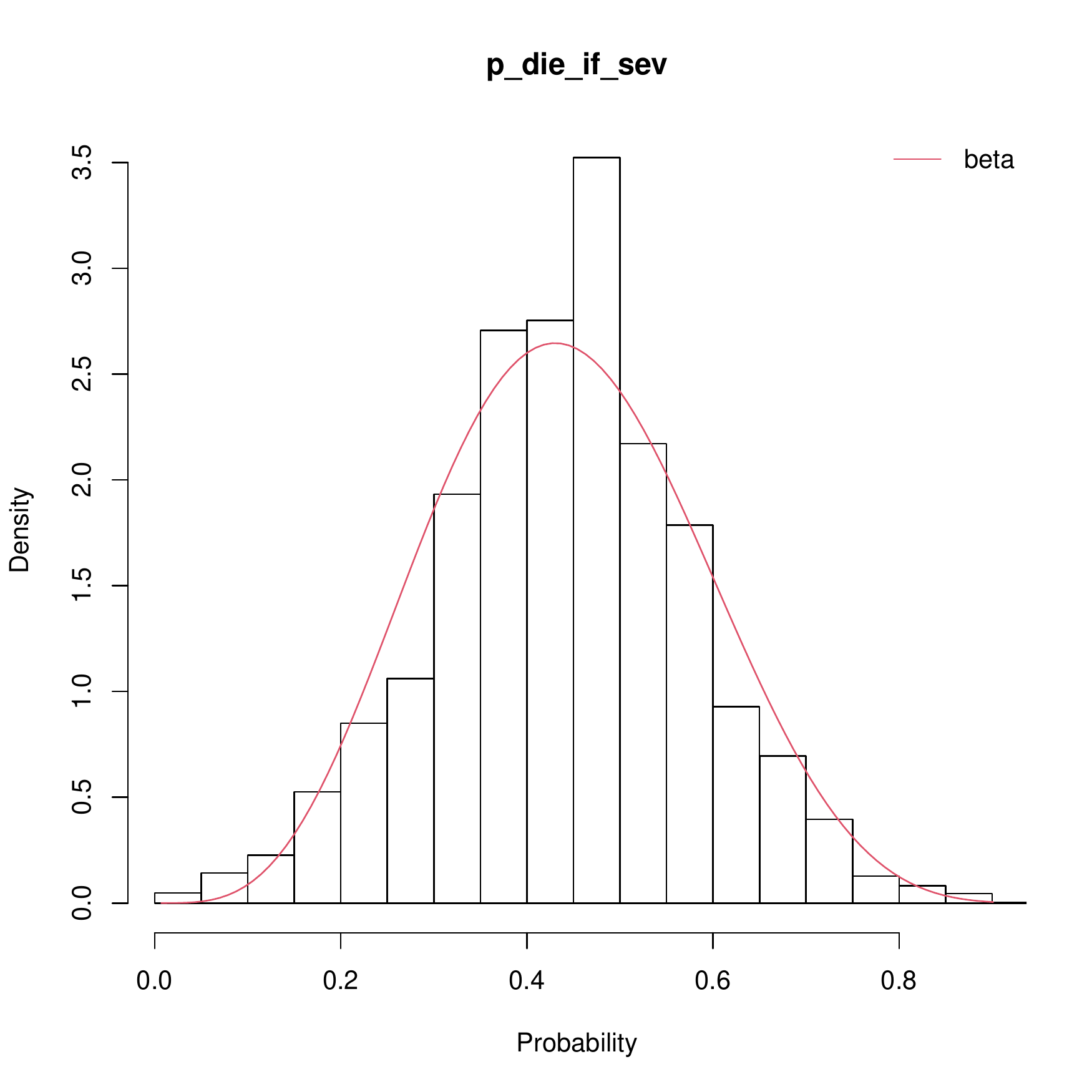}
    \end{tabular}
    \caption{ Data form the Mexican Federal Government open data base consulted on February 24,2020. A) Fitted density function of the times form symptom onset to severe disease   B) Fitted density function of the times from severe symptom onset  to death. C) Fitted density function of the probability of presenting severe symptoms if symptomatic. D) Fitted density function of the probability of dying if showing severe symptoms.}
    \label{fig:dist1}
\end{figure}

During the the epidemic, many generalizations of the basic model SIR or SEIR were developed that included compartments for presymptomatic, symptomatic, asymptomatic, hospitalized, ICU, recovered and dead individuals respectivelly, just to list a few (see e.g.,\cite{MB2020,Li2020,Magal2020,Mena2020,Mukandavire2020,Wu2020,Zhao2020}). Many of the residence times for these compartments, because of the nature of the differential equations of the Kermack-McKendrick models, are assumed to be exponential when in reality they are not. They are usually Beta or Gamma distributions that may also vary as the epidemic evolves \cite{Ali2020}. In Figure \ref{fig:dist1} we show the distributions of  several important variables obtained from the Mexican Federal Government open data \cite{Salud}. These four distribution functions determine in a significant way how individuals are allocated to different model compartments after the transmission event occurs. It is clear that none of them follows the hypothesis of exponentially distributed residence times assumed in Kermack-McKendrick models. Figure \ref{fig:dist1}A shows the empirical distribution and theoretica fit of a Gamma distribution for the time it takes a person to present severe disease once she has shown symptoms. This distribution is fundamental to be able to estimate hospitalization rates.  Figure \ref{fig:dist1}B is the distribution of times from severe symptom onset to death as the epidemic develops, Figure \ref{fig:dist1}C records the probability of presenting severe symptoms if symptomatic. This probability distribution has been obtained for the whole population of Mexico. Obviously, it should be estimated by age and conditional on individuals presenting key comorbidities; finally, Figure \ref{fig:dist1}D presents the probability of dying once showing severe symptoms which is necessary to estimate the expected IFR (infection fatality rate).
One other important distribution, perhaps the most important of them, is the one for the generation and serial interval of the infection. The generation interval is the time that passes between the infection of the infector and the infection of a secondary infectee. It is the fundamental basis for the estimation of the basic reproductive number \cite{Champredon2015,Park2020,Wallinga2007}. The serial interval is an approximation to the generation interval and measures the time beween symptom onset in the infector and symptoms onset in the infectee. A recent study in China gives an estimate of the serial interval is of 5.3 days with an interquartile range of 2.7 to 8.3 days, the generation interval has the same median but the interval of variation is a bit moved to the right, from 3.1 to 8.7 \cite{Sun2021}. When mitigation measures are applied, however, it matters the timing of their application. Still following \cite{Sun2021} the median generation interval increases 4 days when cases are isolated 2 days after symptom onset, and increase to 7 days when the isolation occurs more than 6 days after symptoms onset. Given this variability, it is not surprising the great spread that estimations of the basic reproduction number show. For China, Hunan province outbreak, \cite{Sun2021} gives $R_0=2.19$ while \cite{Sanche2020} gives $R_0=5.7$. This fact highlights the importance of a clear understanding of the underlying hypothesis of the Kermack-McKendrick models, particularly on those that relate to the transmission process, namely $R_0$. The SIR and SEIR models, described in the previous section, render a basic reproductive number of the form $R_0=\beta/\gamma$ which means that the serial or generation interval is approximated by the recovery time  $T_R=1/\gamma$. According to the CDC \cite{CDC2021} the recovery period for COVID-19 in adults with mild disease is 10 days, and for those with severe disease is no longer than 20 days, numbers that are about two or three times larger than the serial interval, implying a possible overestimation of $R_0$ if doing it directly from the Kermack-McKendrick model. 

\begin{figure}[htbp]
	\centering
	\includegraphics[width=110mm]{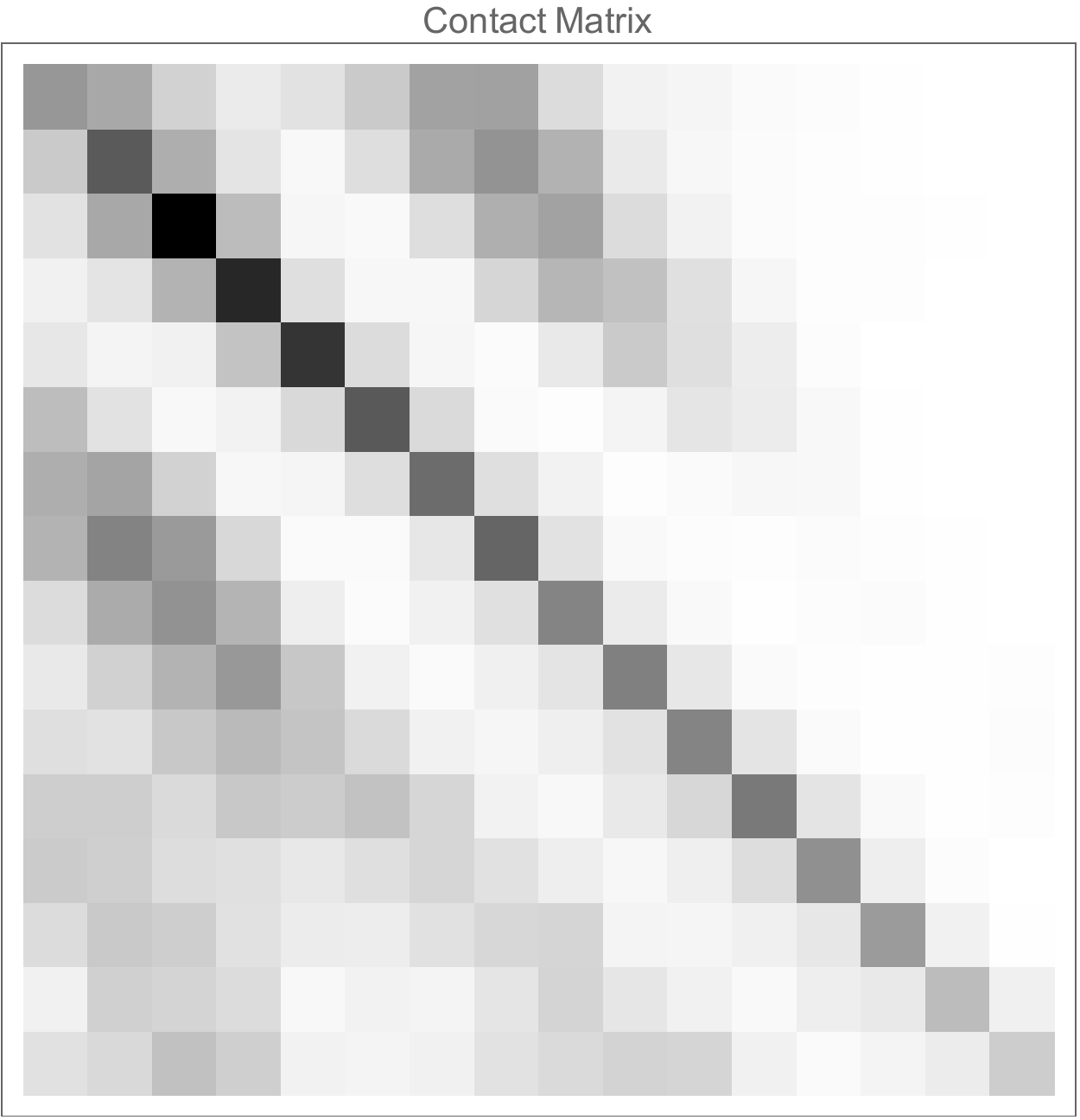}
	\caption{Contact matrix for a home environment modified from \cite{Prem2017}. Each row is an age group of 5 years for a total of 16 age groups.}
	\label{fig:contact}
\end{figure}

Also, in all of these generalized Kermack-McKendrick models, the core of the structure resides in the mixing function used. The mixing function represents the way in which individuals in the population come together and make contact with each other that results in a transmission event. The models in the previous section assume that mixing is random and it is represented by the law of mass action $S\times I$, valid when the population is homogeneous and there are no differences due to age, gender, preferences or any other factor. Define $c_{ij}$ the proportion of contacts that  group $i$ has with group $j$. These contact rates must satisfy the following conditions \cite{Busenberg1991,Glasser2012},:
\begin{enumerate}
    \item $c_{ij}\geq 0$,
    \item $\sum_{j=1}^nc_{ij}=1$ for $j=1,...,n.$,
    \item $a_iN_ic_{ij}=a_jN_jc_{ji}$,
\end{enumerate}
where $N_i$ is the size of group $i$ and $a_i$ is the activity or risk level of group $i$. Proportional mixing between groups $i$ and $j$ can be obtained by assuming that $c_{ij}=c^a_ic^b_j$ and then axioms 2 and 3 above render $c_{ij}=Cc_j^b=\hat{c}_j$ with
$$\hat{c}_j=\frac{a_jN_j}{\sum_{i=1}^na_iN_i}.$$
Preferential mixing assumes that contacts of a given group $i$ consists of reserving a fraction $\epsilon_i$ of contacts for within group contacts (group $i$) and a fraction $1-\epsilon_i$, is distributed among all other groups according to 
\begin{equation}\label{eq:pref}
    c_{ij}=\epsilon_i\delta_{ij}+(1-\epsilon_i)\frac{(1-\epsilon_j)a_jN_j}{\sum_{i=1}^n(1-\epsilon_j)a_iN_i},
\end{equation}
with $\delta_{ij}=0$ if $i\neq j$ and $\delta_{ij}=1$ if $i= j.$ 
Heterogeneity is a property of human populations. The mass action mixing function clearly oversimplifies the way groups of individuals interact. It is true that it can render reasonable approximations at a very general level but one has to be cautious in this respect. Many mathematical models for COVID-19 include several types of infectious individuals including presymptomatic ones, that  still follow the mass action law for their contacts allowing only for a reduction or increase in the probability of infection. 
In Figure \ref{fig:contact} we show a typical contact matrix obtained from an extensive study on influenza like illnesses \cite{Prem2017}. The clear diagonally dominant pattern is shown. 
The use of contact matrices is of fundamental importance when designing protocols for ending the lockdowns or vaccination strategies. Age is a natural risk factor in this disease and is one that can be directly incorporated into the Kermack-McKendrick framework. As a representative example, we follow the model in \cite{Zhao2020}. This model was constructed to help in the design of an strategy to reopen the economy after the ending of strict lockdown in many regions of the world. Risk of contagion has differential characteristics depending on age. It is known, for example, that the IFR (infection fatality rate) is higher in older people (60 yrs old and older). So, release of older adults at the same time than younger individuals may increase the IFR because of higher transmission and mortality in the older adults age group \cite{Bubar2020}. In order to design sound strategies, models have to include age as a new variable and, more importantly, consider that because of the mitigation measures, the structure of pre-pandemic contact patterns most likely have  changed. The model \cite{Zhao2020} introduces seven compartments for each age class considered which are $S_i$ susceptible, $E_i$ exposed, $A_i$ asymptomatically infected, $I_i$ symptomatically infected, $H_i$ hospitalized, $R_i$ recovered and $M$ dead individuals where $i$ refers to the age class. There are three age classes: 0-44 years old, 45-64 years old and older than 65 years old. For each age class the parameters are $1/k_i$ mean latent period, $1/\gamma_i$, $1/\gamma_{ia}$ the mean infectious period for symptomatica and asymptomatic individuals respectivelly, $\eta_i$ the average rate of transition form $I_i$ to $H_i$, $\phi_i$ the rate ay which individuals abandon the $H_i$ compartment, where $q_i$ is the proportion of theses that die, $p_i$ is the proportion of infections that develop symptoms, and $\theta_i$ and $\xi_i$ denote the proportional reduction of infectivity, with respect of symptomatic cases, of asymptomatic and hospitalized individuals of age class $i$.
The force of infection for each age class in this model incorporates the contact structure:
$$\lambda_i(t)=\sum_{j=1}^3a_i(t)c_{ij}\frac{I_j+\theta_jA_j+\xi_jH_j}{N_j}$$
where $c_{ij}$ is defined as in (\ref{eq:pref}). The equations are
\begin{align}
    S_i'&=-S_i\lambda_i(t),\\\nonumber
    E_i'&=S_i\lambda_i(t)-k_iE_i,\\\nonumber
    A_i'&=(1-p_i)k_iE_i-\gamma_{ia}A_i,\\\nonumber
    I_i'&=p_ik_iE_i-(\gamma_i+\eta_i+\delta_i)I_i,\\\nonumber
    H_i'&=\eta_iI_i-\phi_iH_i,\\\nonumber
    R_i'&=\gamma_{ia}A_i+\gamma_iI_i+(1-q_i)\phi_iH_i,\\\nonumber
    M_i'&=q_i\phi_iH_i+\delta_iI_i,\nonumber
\end{align}
where $i=1,2,3$ denote the age classes previuosly discussed. The main conclusion of this work is ``...that the early release of low-risk young [...] may protect older and more vulnerable individuals from severe health risks. As an added benefit, the staggered-release policy allows for more activity in the population at an earlier date than the delayed simultaneous-release policy would. Thus staggered-release policies can positively impact everyone involved by allowing people to return to their lives without risking them" \cite{Zhao2020}.

\section{Modification of the basic models}
One of the important measures taken to control de spread of the epidemic, was the application of mitigation measures expressed in total or partial lockdowns, social distancing, use of facemasks and frequent washing of hands among others. We will review here a model developed in \cite{Santana2020b} for a partial lockdown. The epidemiological model consists of the compartments of susceptible, exposed, asymptomatically infected, symptomatically infected, reported, recovered and death individuals. It is assumed that the population is split into two different subpopulations one that fully complies with social distancing, use of face masks, mobility restrictions, etc., and another group with a lower compliance. These different levels of compliance affect the reduction of the effective contact rate under lockdown. We also consider that individuals under lockdown get tired of it and start to have high risk behaviours that equal their lower contact rates with those of the group with lower compliance. An important parameter in this model is $\omega(t)$ the variable rate of mobility of the population in lockdown. We now present the model where the index $i=1,2$ indicate the two groups $i=1$ high compliance, $i=2$ lower compliance:

\begin{equation}\label{eq:split}
	\begin{aligned}
		S_{i}' &= \mu_{h}\left(S_{i} + E_{i} + I_{ai} + I_{si}\right) + \left[\left(2-i\right)q + \left(i-1\right)\left(1-q\right)\right]\mu_{h}\left(I_{r} + R\right)\\
		&\ \ \ - \beta_{i}(t) \left({I_{si}} + {I_{ai}}\right)\frac{S_{i}}{N^*} + (-1)^{i}\omega(t)S_{1} - \mu_{h}S_{i} + \left(1-i\right)^{i}\chi_{r}R\\
		E_{i}' &= \beta_{i}(t) \left({I_{si}} + {I_{ai}}\right)\frac{S_{i}}{N^*} - \gamma E_{i} + (-1)^{i}\omega(t)E_{1} - \mu_{h}E_{i}\\
		{I_{ai}}' &= \rho\gamma E_{i} - \eta_a I_{ai} + (-1)^{i}\omega(t)I_{a1} - \mu_{h}I_{ai}\\
		{I_{si}}' &= (1 - \rho)\gamma E_{i} - \left(\eta_{s} + \delta_{s}\right)I_{si} + (-1)^{i}\omega(t)I_{s1} - \mu_{h}I_{si}\\
		{I_{r}}' &=  \delta_{s}\left(I_{s1} + I_{s2}\right) - \delta_{r}I_{r} - \mu_{h}I_{r}\\
		R' &= \eta_{a}\left(I_{a1} + I_{a2}\right) + \eta_{s}\left(I_{s1} + I_{s2}\right) + \left(1 - \mu\right) \delta_{r}I_{r} - \mu_{h}R - \chi_{r}R\\
		D' &= \mu \delta_{r}I_{r}
	\end{aligned}
\end{equation}
where $N^* = \sum_{i=1}^{2}\left(S_{i} + E_{i} + I_{ai} + I_{si}\right) + R$, $i=1,2$; $\mu_h$, is the birth rate and natural death rate of the population, and $\chi_r^{-1}$, the average time of temporal immunity. Individuals can abandon the group with higher compliance at a rate $\omega$,
$\gamma^{-1}$ is the incubation period , $\rho$ is the proportion of individuals that become asymptomatically infected, $\eta_a^{-1}$ is the average recovery time for asymptomatic individuals, $\eta_s^{-1}$ is the average recovery  time for symptomatic individuals, $\delta_s^{-1}$is the average time that takes one individual to seek medical attention, $\delta_r^{-1}$ is the average time until recovery and $\mu$ is the proportion of reported individuals that die.

The effective contact  rates for each group, $\beta_{i}(t)$, decrease after the implementation of the lockdown but is not the same for each subpopulation. There are many alternatives to model this behavior. One way is to define the following:
\begin{align}\label{eq.beta_step}
\beta_i(t) = \begin{cases}
bq_1, & \text{if } t<T;\\
bq_2, & \text{if } t\geq T,
\end{cases}
\end{align}
where  $q_1, q_2 \in [0,1]$ and $(1-q_i)\times100$ are the percentage reduction of the original (pre-lockdown) contact rate $b$, is a step function.

This model allows us to explore the changes in the epidemic curve have when the effective contact rate has a sudden increase because of, for example, short burst of hightened mobility. Increases in the effective contact rate are related to human behavior. The reasons behind violating the rules of lockdowns are very varied and very rarely predictable except perhaps, when holidays, vacations of other festivities that have a social component of family, religion or political nature, occur. These atypical events during the epidemic can be characterizes as superspreading events (events with hightened transmission in alarge scale) that, in the context of the model in this section, we associate them with the rate $\omega$ that represents the losening of compliance of lockdown because individuals in lockdown move out of it and become, de facto,  members of the low compliance group.

An atypical increase in mobility is modeled as follows \cite{Santana2021}:
First, the increase in mobility lasts only for a period of $\tau$ days; second, the change in mobility is reflected on $\omega_0$ by a factor $k$, i.e., for the superspreading period, the new compliance-failure rate is $k\omega_0$ the baseline mobility rate; third, behavioral change is not instantaneous but modelled as
	 \begin{align}\label{Eq:beta1}
	 \beta_i(t) = 
	 \begin{cases}
	 b - \frac{(1 - q_i)}{\theta}b\left(t - T\right), & T\leq t < T+\theta,\\
	 q_i b, &  t \geq T+\theta,
	 \end{cases}
	 \end{align}
where $\theta$ is the period where the contact rate declines, and $q_{i}$ is the proportion of reduction of the baseline effective contact rate for each subpopulation. 

\begin{figure}[htbp]
	\centering
	\includegraphics[width=12cm]{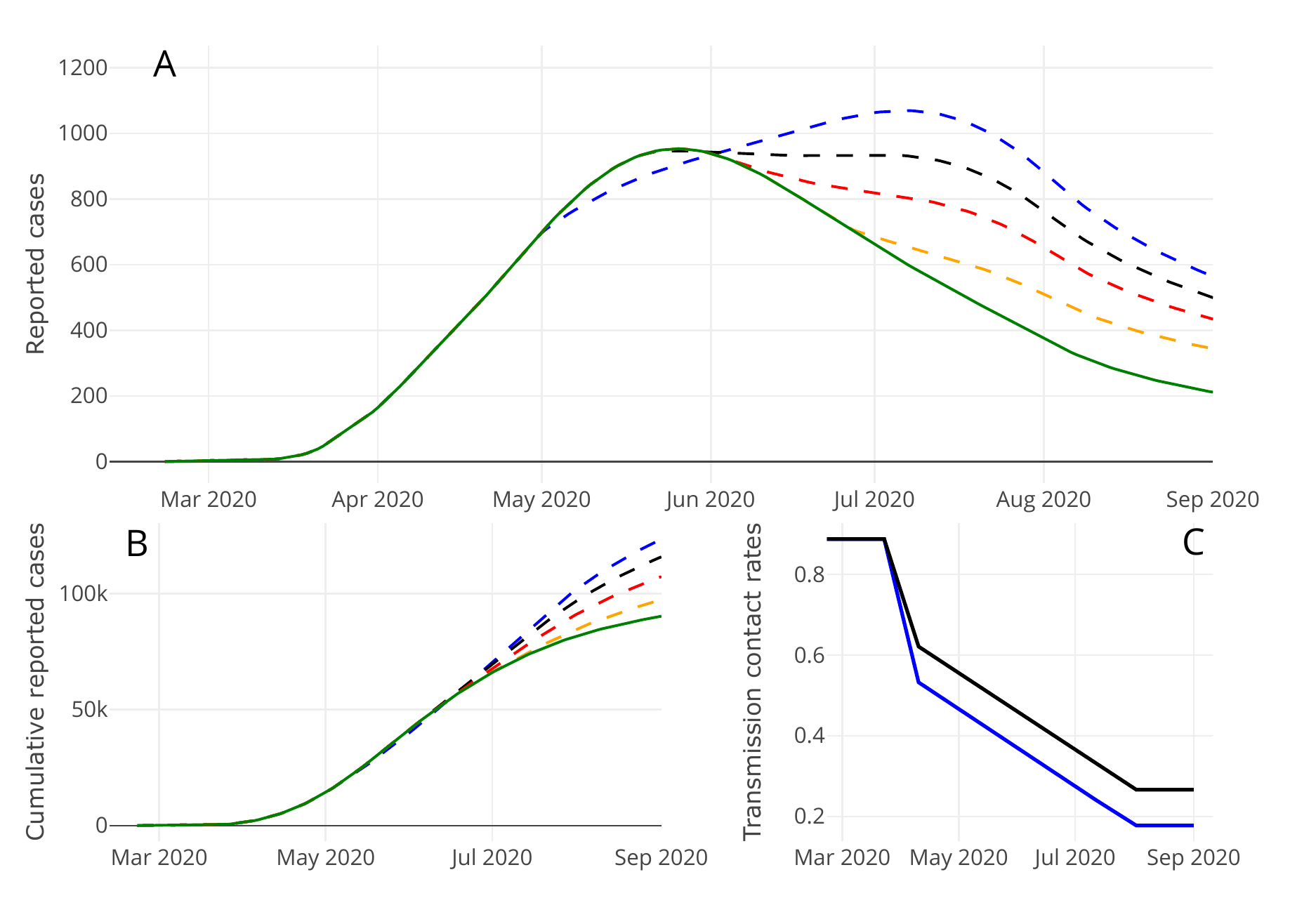}
	\caption{Impact of different times of mobility increase on the epidemic curve. Mobility increase is given by $10 \omega_0$ $(\omega_0 = 0.005)$. A) The green line is the baseline epidemic curve. Blue, black, red, and yellow discontinuous lines illustrate the scenarios when the  mobility event starts four weeks before, a week before, a week after, and four weeks after peak incidence, respectively. B) Cummulative incidence, C) effective contact rates. }
	\label{Figure10}
\end{figure}

As an example of the results we assume no reinfections occur ($\chi_{r} = 0$) and  simulate one atypical event where mobility increases for three days by a factor of $k=10$ times the baseline failure rate, i.e. $\omega = 10\omega_0$, with $\omega_0 = 0.005$). We simulate several events before or after the baseline incidence peak. In Figure \ref{Figure10} the green solid line is the baseline epidemic curve, normal Kermack-McKendrick dynamics. Note that the worst-case scenario (blue dashed line)  occurs when the atypical event is located on the exponential growth phase of the curve and relatively far from the maximum incidence. The most beningn scenario occurs when the superspreading event happens after the incidence peak (dashed yellow line). When the superspreading event occurs just before (black dashed line) or just after (red dashed line) the peak of the curve, a plateau that lasts several weeks is formed. Figure \ref{Figure10}B are the cumulative incidence for each case and shows that, regardless of the timing of  increased mobility, there is a similar increment in cases. Figure \ref{Figure10}C illustrates the shape of the effective contact rates. Further details can be found in \cite{Santana2021}
This model was applied to the epidemic in Mexico City. One of the important conclusions was the accurate prediction of time when maximum incidence would be reached that was located towards the end of May. With this model the entrance of the epidemic to a long plateau was also correctly predicted. 

\section{Epidemic curves}
\begin{figure}
    \begin{tabular}{cc}
         A&B\\\includegraphics[width=6cm]{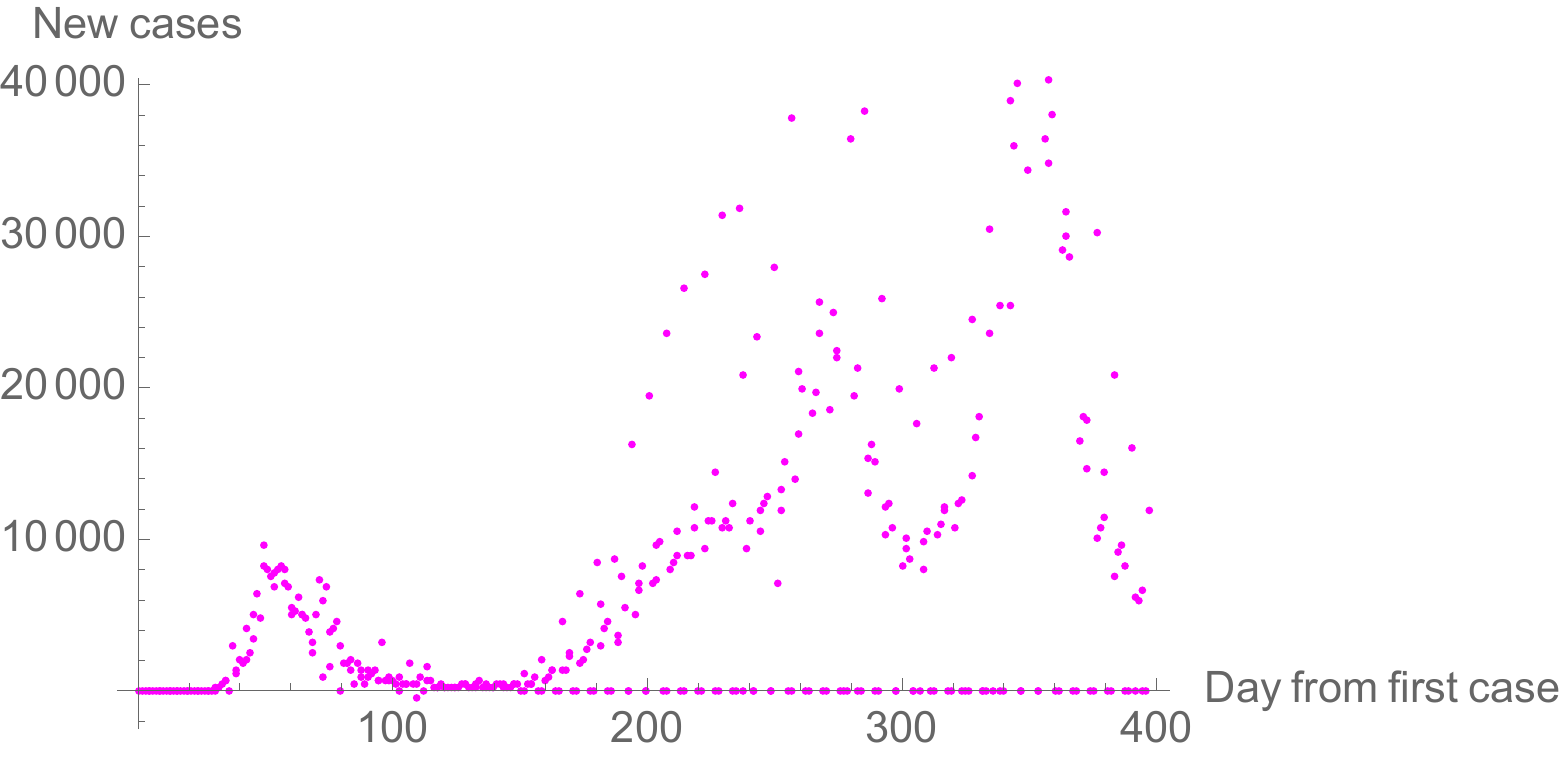}& \includegraphics[width=6cm]{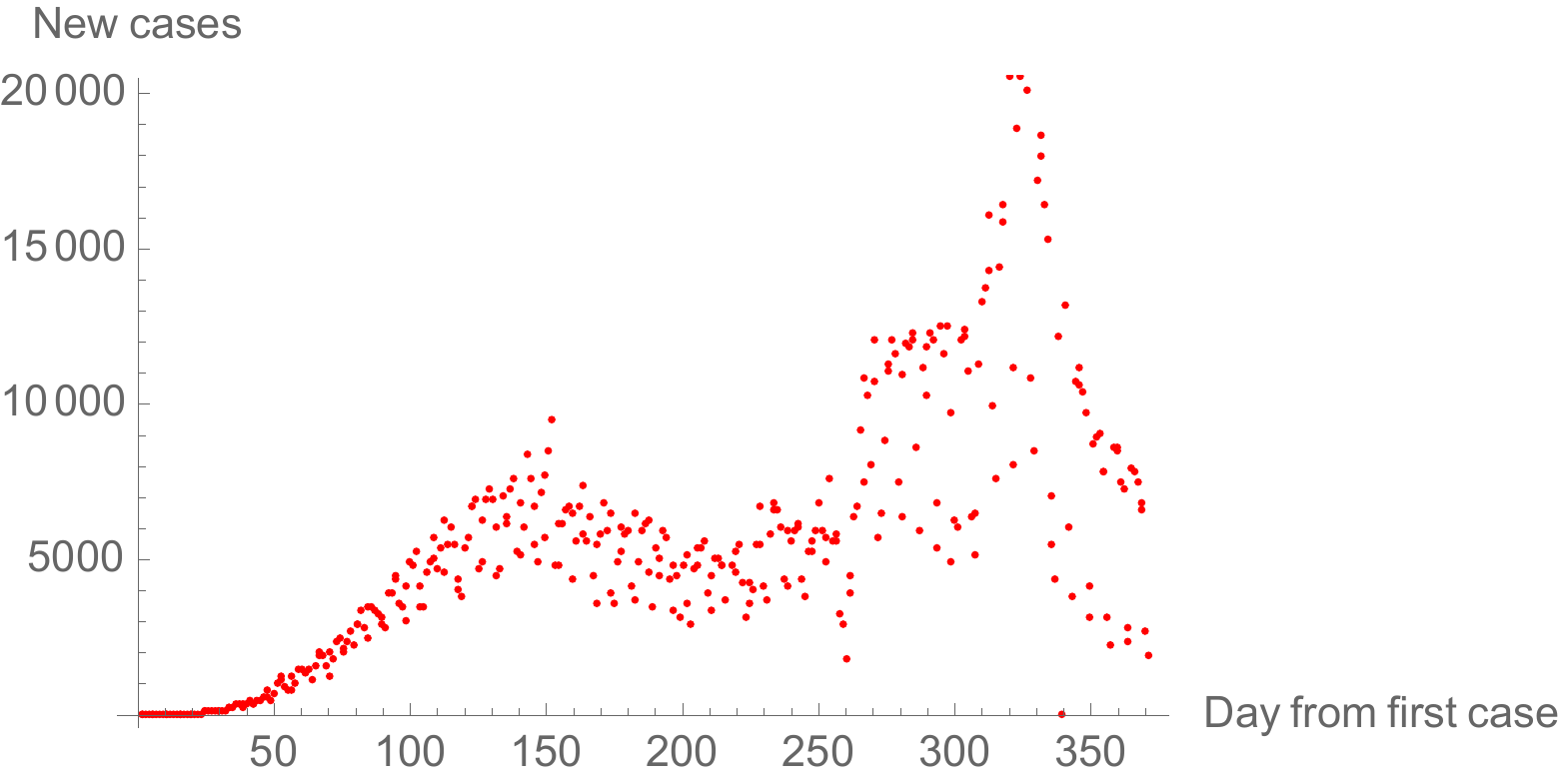}  \\
         C&D\\
         \includegraphics[width=6cm]{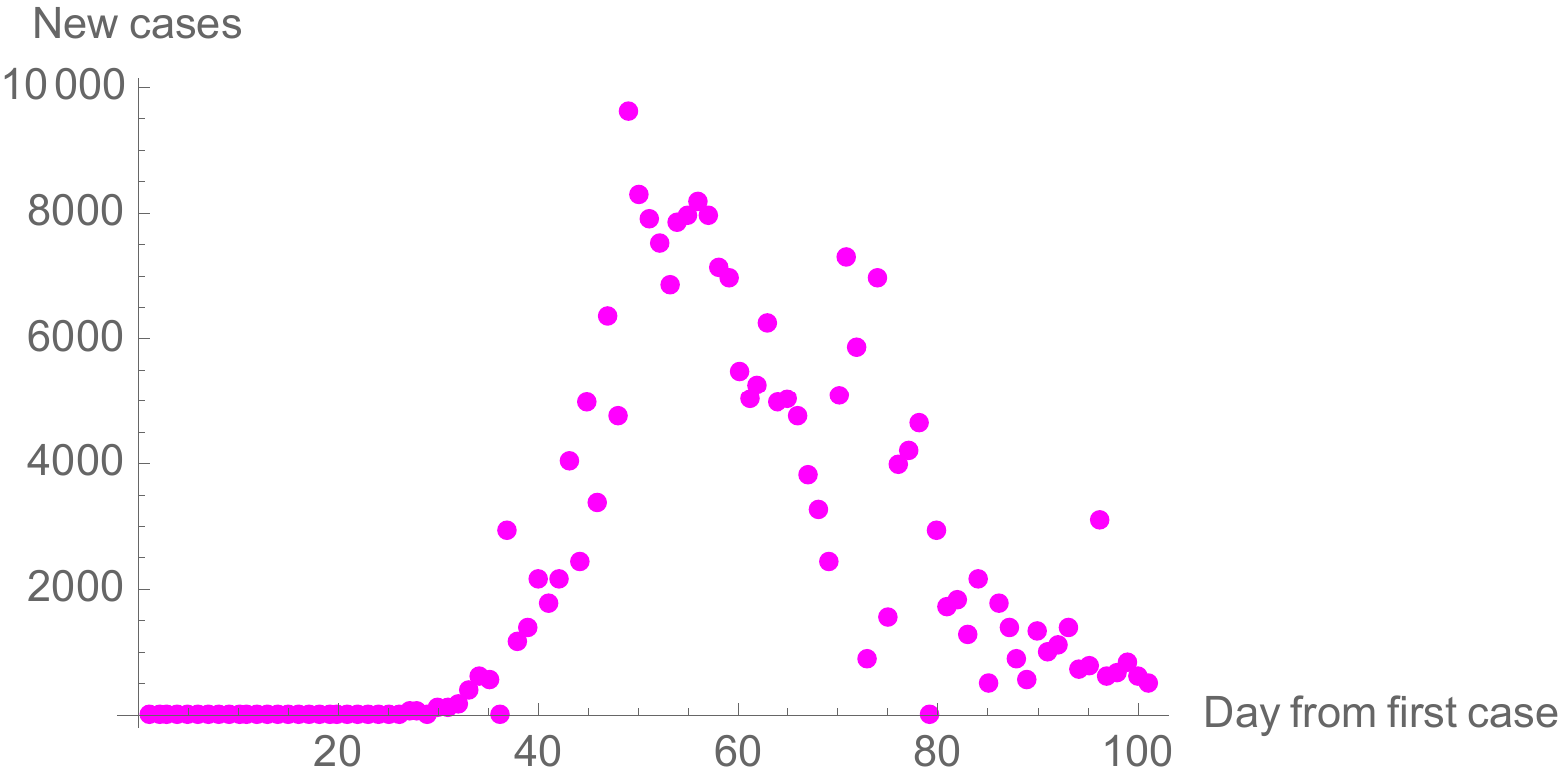} & \includegraphics[width=6cm]{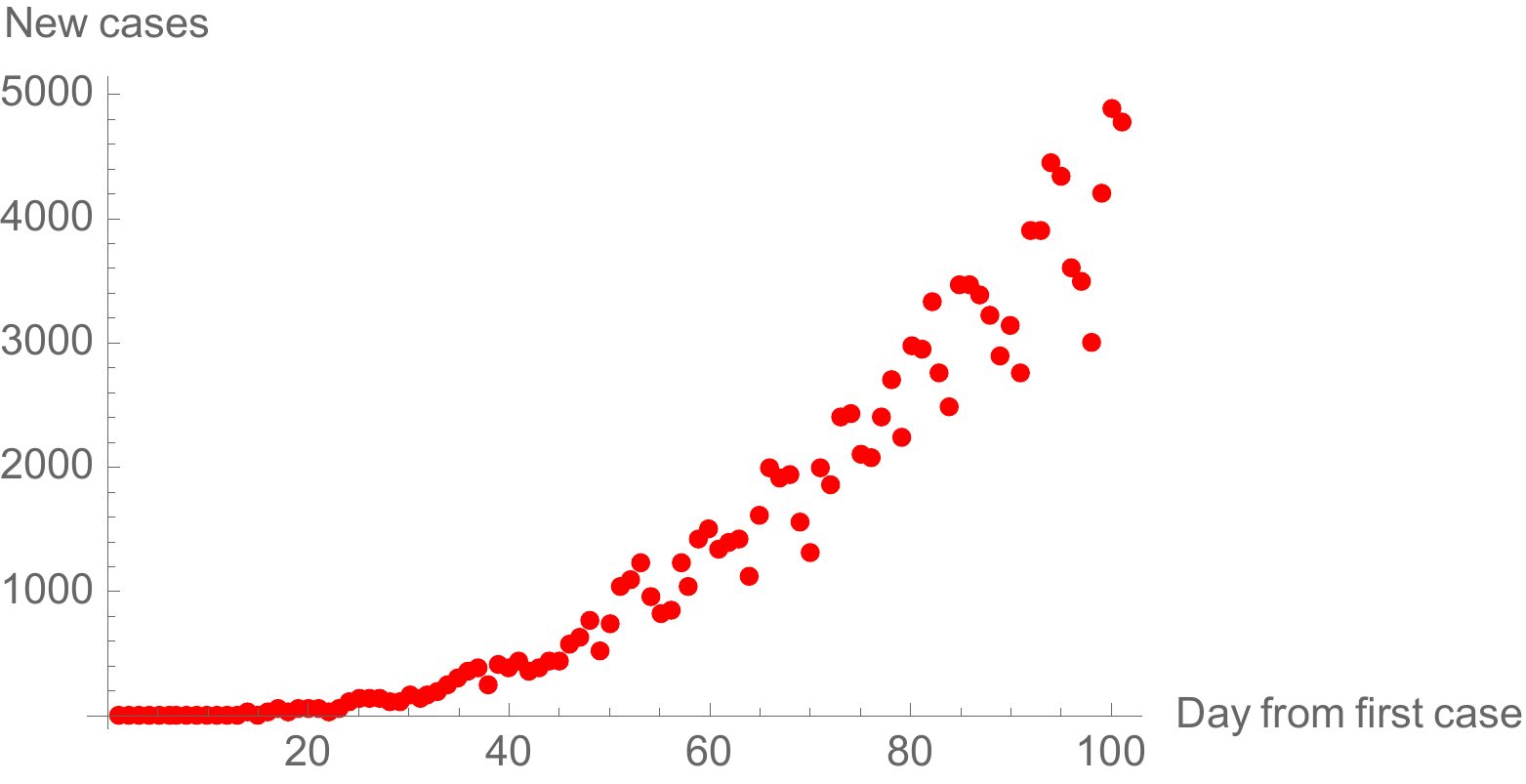}
    \end{tabular}
    \caption{Comparative plots of the epidemic curves for Mexico and Spain. A)and B) Epidemic curve of Spain and Mexico, respectivelly, from the first case to early March 2021; C) and D) initial epidemic growth of Spain and Mexico, respectivelly, up to day 100 of their respective epidemics. Data from the Our World in Data web page}
    \label{fig:SpainMex}
\end{figure}

\begin{figure}
    \begin{tabular}{cc}
         A&B\\\includegraphics[width=6.5cm]{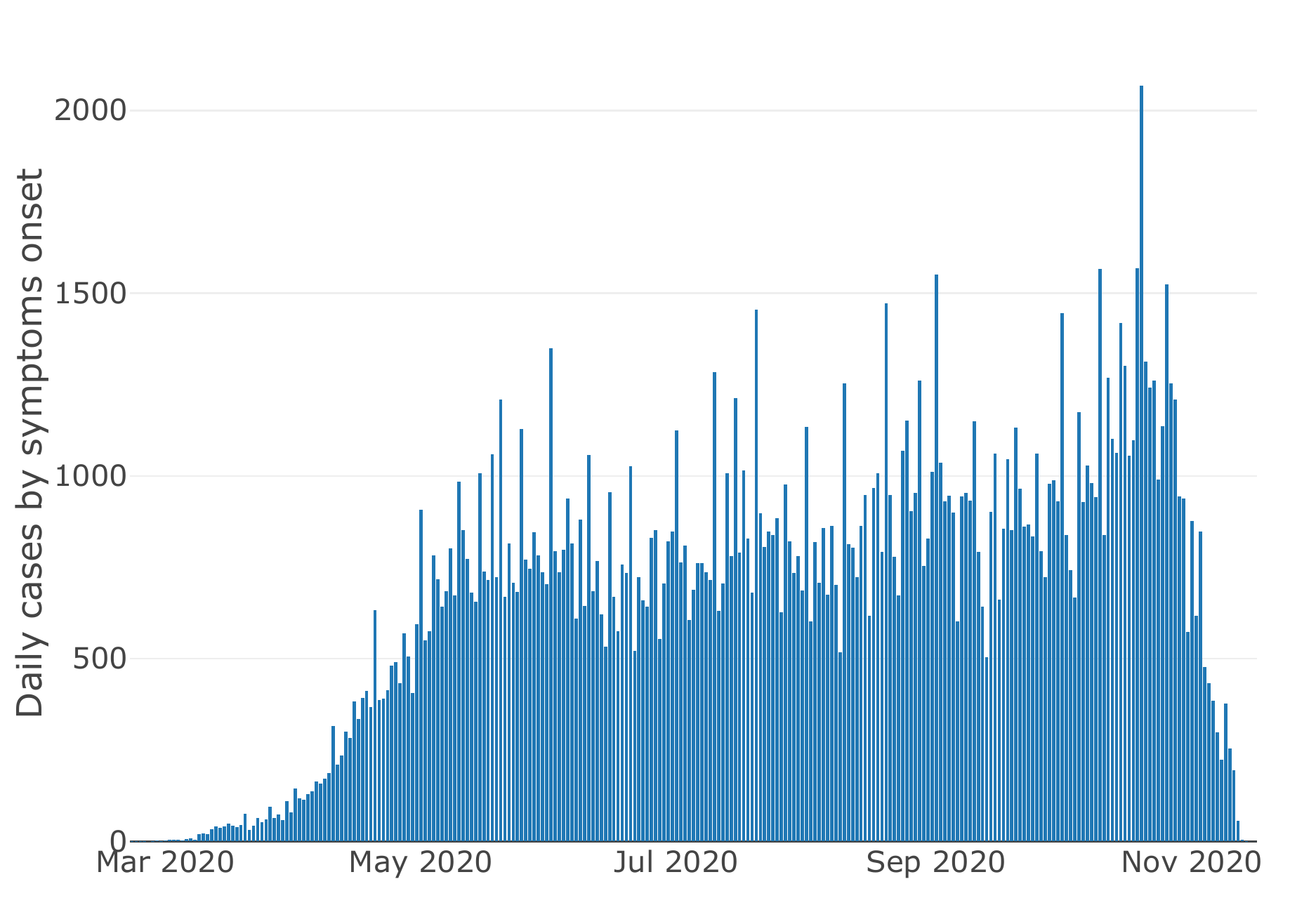}& \includegraphics[width=6.5cm]{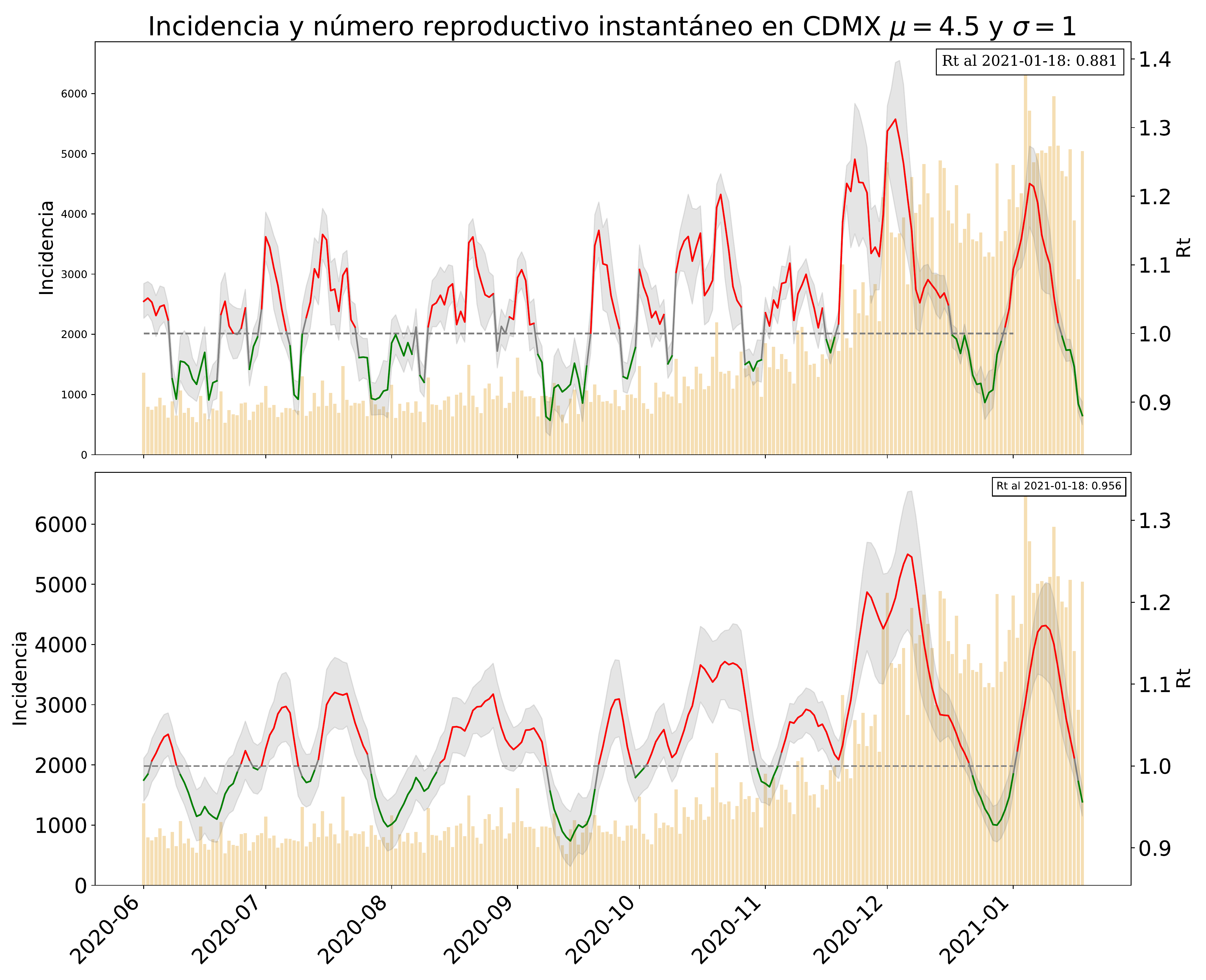}  
    \end{tabular}
    \caption{A) Epidemic curve for Mexico City. The initial linear growth and the plateau reached after May 30th can be identified until about mid September, B) Evolution of the instantaneous reproductive number $R_t$ for México City from June 1 2020 to the end of February 2021. The erratic oscillations around $R_t=1$ are clear. The lower panel in B shows an smoothed version of $R_t$.}
  \label{fig:CDMX}
\end{figure}

Kermack-McKendrick models have, all of them, an initial exponential growth and then, after reaching a maximum, the epidemic curve decreases also exponentially. In Figure \ref{fig:SpainMex} the epidemic curves of Spain and Mexico, from the initial case up to March 9, are plotted. 
The differences of both curves are striking. Spain reached the epidemic peak very fast (around day 50) reaching almost 10 000 cases per day at the acme and then, because of the strongly enforced lockdown measures, the first epidemic outbreak practicaly died out (Figure \ref{fig:SpainMex}A and C); in contrast, the epidemic in Mexico grew very slowly likewise reaching almost the 10 000 cases per day but until the 100th day of the epidemic; then because of the very weak mitigation measures, the epidemic reached a plateau and stay there for several months until mid september where the second major outbreak started (Figure \ref{fig:SpainMex}B and D). The initial growth of the Spanish curve is compatible with exponential growth while the Mexican curve grows practically in a linear way. This peculiar behavior observed in many epidemic growth curves around the world, has been explored  in \cite{Thurner2020} who have provided an explanation for the this linear growth that involves the structure of the underlying contact network of individuals and the maintenance of the effective reproduction number around $R_t\approx 1$ for sustained periods of time. This phenomenon can be better appreciated in the epidemic of Mexico City (Figure \ref{fig:CDMX}).

From the results of model (\ref{eq:split}) and the mitigation measures applied in the Mexico City epidemic, we can deduce that the initial slow growth
was the product of an early application of the weak mitigation measures. Indeed the first case in Mexico occurred by the end of February 2020 and the first mitigation measures where applied on March 23th about 25 days after the initial case. From \cite{MB2020,Santana2020b} and the results of model (\ref{eq:split}) discussed in the previous section, we can conclude that there were superdispersion events in Mexico City in early May (particularly April 30th to May 10th) shifted the expected day of maximum incidence towards the end of the month and also sent the epidemic into a quasi-stationary state characterized by values of $R_t\approx 1$. 
In contrast, in Spain the first case was recorded by the end of January and the lockdown was imposed on March 14 more than 40 days after the initial case. Given the stricter nature of the lockdown, the initial outbreak was controlled and resolved by the 100th day of the epidemic.

\section{Epidemic interactions}
Vaccines deployment to protect against severe acute respiratory infections is a priority around the globe as the second wave of Covid-19  still grows in many parts of the world and there is increasing pressure on health-care systems to accomodate the growing numbers of severe cases. In addition to this, acute respiratory 
infections occur as annual or biannual epidemics as is the case for influenza.
With the presence of  SARS-CoV-2 we are faced with an scenario were this virus will cocirculate with the influenza viruses the whole year but particularly in the fall and winter seasons in each hemisphere. Fortunatelly, there are vaccines for influenza and therefore, influenza vaccination can potentially reduce its prevalence and reduce symptoms that might be confused with those of Covid-19 \cite{Sultana2020}.
In \cite{Acuna2021} a mathematical model is presented for the cocirculation of SARS-CoV-2 and influenza under the assumption that a vaccine exists for both viruses. The model explores the effects of coverage, efficacy, induced temporal immunity on the dynamics of this coupled system.

$$\begin{aligned}
		s' &= \mu  - \left(\beta_i(t)(i+i_{r_y})+ \beta_y(t)(y +y_{r_i})\right)s+
		\theta_ir_i+\theta_yr_y+\theta r -(\mu +\phi_i+\phi_y)s,  \\
		i' &= \beta_i(t)(i + i_{r_y})s - \alpha\beta_y(t)iy - (\eta_i  + \mu) i, \\
		y' &= \beta_y(t)(y + y_{r_i})s + \alpha \beta_y(t)iy - (\eta_y +\mu )y, \\
		r_i' &= \phi_i s + \eta_ii-(1 - p_{i})\beta_y(t)r_i(y+y_{ri})-(\phi_y + \theta_i+\mu)r_i,\\
		r_y'&= \phi_y s + \eta_y y-\beta_i(t)r_y(i+i_{ry})-(\phi_i +\theta_y+\mu)r_y, \\
		y_{r_i}' &= (1 - p_{i})\beta_y(t)r_i(y+y_{ri}) - (\omega_y + \mu)y_{r_i}, \\
		i_{r_y}' &= \beta_i(t)r_y(i+i_{ry}) - (\omega_i + \mu)i_{r_y},\\
		r' &= \omega_y y_{r_i} + \omega_i i_{r_y} + \phi_y r_i + \phi_i r_y - (\theta + \mu)r, 
\end{aligned}
$$

\begin{figure}[htbp]
	\centering
	\includegraphics[width=120mm]{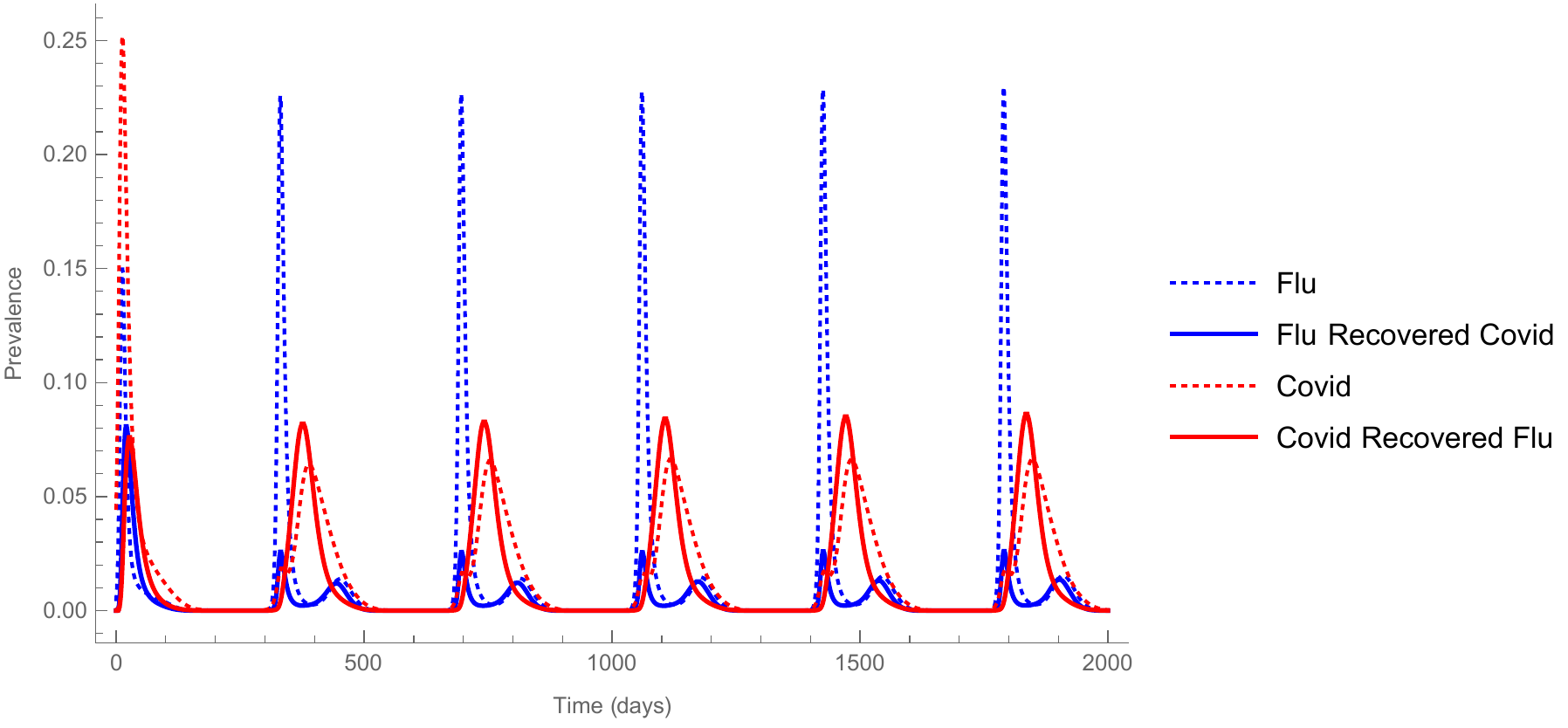}
	\caption{Outbreaks of coronavirus and influenza undergoing alternate occurrences both for primary and secondary infections.Shown are primary coronavirus cases (dashed red), secondary coronavirus cases (solid red), primary influenza cases (dashed blue lines) and secondary influenza cases (solid blue). }
	\label{fig:inco}
\end{figure}

\medskip\noindent
where $s$, $i$, $y$, $r_i$, $r_y$, $y_{ri}$, $i_{ry}$, $r$ represent, respectively, the individuals that are susceptible, infected by influenza, infected by coronavirus, immune against  influenza, immune againts coronavirus, infected by coronavirus while immune to influenza, infected by influenza while immune to coronavirus and immune for both diseases. The model couples two SIR models where vaccinated and naturally immune individuals for a given disease (i.e., individuals that are immune after recovering from an infection) are pooled together in the same compartments); $\eta_k$,  $\omega_k$, $\theta_k$,  $\theta$, $\phi_k$ are the recovery from a primary infection, recovery from a secondary infection, waning from a primary infection/vaccination, waning from a secondary infection/vaccination, and vaccination rates for influenza $(k=i)$ or coronavirus ($k=y$), respectivelly. Finally $\beta_k(t)$ are the time dependent effective contact rates and $\alpha$ is a coefficient that introduces the hypothesis that a host already infected by the influenza virus can be infected by the coronavirus resulting in the effective displacement of influenza from that host. As mentioned earlier, the flu season of 2019-2020 has been atypical.There has been a significant decrease of influenza cases in the season that, at the time of writing, is about to end, and also SARS-CoV-2 has been the dominat respiratory disease from the beginning of 2020 \cite{WHO2020}. There are several issues dealt with in the paper where this model was presented \cite{Acuna2021} that have to do with the indirect protective effect of the influenza vaccine against coronavirus infection, and the consequences of deficiencies in coverage and vaccine efficacy on the persistance of one or both diseases. We concentrate here in a particular phenomenon of the epidemic. At the time of writing the World is in the midst of a vaccination campaign aim to control epidemic growth by decreasing the risk of infection which would allow everybody to return to a new normality. The question however, is if the coronavirus will really go away. The basic SIR and SEIR models presented in the first section of this work can give us plausible scenarios and clues to consider and evaluate. To study the long-term behavior of both epidemics we need to assume that the effective contact rates are time dependent. Take $\beta_k(t)=a_k (1+\cos(2 \pi t/365))$ which is a periodic function with  period equal to 365 days. The question is then if the coronavirus persists as an endemic disease, what kind of dynamics will it have? Will it become similar to influenza with annual or biannual epidemic peaks? Will influenza maintain the same trend as the one observed until 2019? The results of some of the simulations are shown in Figure \ref{fig:inco}. It can be seen that influenza and the coronavirus can present joint yearly epidemic episodes. Other possibilities for epidemic episodes are explored in \cite{Acuna2021}.

Influenza and influenza-like illnesses cocirculating with the coronavirus constitute a single group of infectious agents that is interacting with SARS-CoV-2. The large disruptions that lockdowns and other mitigation measures caused in the world, impacted control programs for many other diseases. Vaccination campains were disrupted for measles, nebulization and mosquito control programs against Dengue. Zika, Chikungunya, Malaria also were severely disrupted. Models are needed to evaluate the current condition of these diseases and to design adequate ways of reduce the morbidity that the post-coronavirus emergency will bring.

\section{Conclusions}
The mathematical modelling of infectious diseases took center stage during the past  year. Many of the models that are being used to inform public health policies are of the Kermak-McKendrick type constructed in multidisciplinary teams of many types of professionals but where epidemiologists and medical practitioners play an important role as experts in the field.  In one of the latest and interesting (and useful) applications we find two works \cite{Bubar2020,Subramanian2020}. In \cite{Bubar2020} several mathematical models are used to explore strategies that might be optimal for  vaccine application now that vaccines are scarce in many regions of the world. The conclusions are that prioritizing age groups between 20 and 49 years old, minimizes cumulative incidence but targeting age classes 60 years and older, mortality and years of life lost are minimized. The authors conclude their work pointing out the multifacetic  problem of deciding which group is first in receiving the vaccine. There are, it is said, also ethical questions that must be takien into consideration. COVID-19 burden has been allocated much as it has been in other pandemics, on certain unprivileged and economically challenged groups of people.  The prioritization derived from the analysis of these models where the main risk-defining quality is age, have to be complemented with other measures of risk. Mathematical models are essentially tools supported by the hypothesis that researchers introduce to construct them. Age is not the only risk variable that can be considered. In Latinamerican countries the comorbidities (diabetes, hypetension, obesity) have increased the fatality rate in younger people and social and economical inequalities threaten the roll out of vaccines. These factors need to be integrated at regional and local levels to evaluate in a more comprehensive way the optimal prirotization of vaccine applications 

The other paper \cite{Subramanian2020}, addresses the problem of the impact on transmission of the large pool of asymptomatically infected individuals characteristic of SARS-CoV-2. In these set of models another important compartment is introduced, that of presymptomatic individuals, and a basic public health tool, worrisomely scarce in many Latinamerican countries: testing capacity and strategies. 

Several issues have not been addressed in this chapter, several of which  represent opportunities for research in areas related to quantitative epidemiology: the role of heterogeneities in the transmission of SARS-CoV-2 \cite{Sun2021}, the need to perform prevalence estimation to monitor the true behaviour of the epidemic \cite{Clark2021},the role of testing as a way to successful screening of populations \cite{Larremore2021}, the dynamics of whithin-host viral interactions \cite{Hernandez2020}.

Finally, a few comments on the use of models in this pandemic. Public opinion on the usefulness and value of models have shifted during this pandemic (e.g, \cite{Kreps2020}) going from a very positive perspective in the initial months, to an esceptical and sometimes disqualifying attitude in recent times. Many Kermack-McKndrick variants were used in the early times of the emergency to forecast its development and many of them failed when the biological, social and behavioral evolving nature of the epidemic started to act. Experts were not that surprised by this development but many modelers and politicians were indeed. Epidemiological models in pandemic times are immensely valuble tools that, as any tool, require careful design and application by multidisciplinary teams and interdisciplinary approaches, to be effective. They also requiere adequate public communcation and dissemination of results.  The mathematical models included in this work are meant to illustrate their use for so-called strategic objectives. In these models the aim is not to fit curves with 95\% confidence intervals and then extrapolate the incidence to see when the epidmic will end, but to understand the basic mechanisms of the problem and provide scenarios supported by current epidemiological and medical knowledge on the pandemic. They are examples, chosen subjectivelly, because they are interesting and informative. No more than that. I hope to have transmitted after reading this chapter, that strategic mathematical models are powerful conceptual ideas that can inform in a very effective way public health decisions. Model fitting is but one component of mathematical modelling whose usefulness is conditioned on a conceptually sound and well thought mathematical model.

\section*{Acknowlegments} I am very grateful to the enjoyable and fruitful collaborations that started early last year and have continued into the present time, with my collegues Adrián Acuña Zegarra, Mario Santana Cibrian, Mayra Nuñez Lopez, Fernando Saldaña García and Ruth Corona-Moreno. Also, I acknowledge support from DGAPA-PAPIIT-UNAM grant IV100220 (convocatoria especial COVID-19) and DGAPA-PAPIIT-UNAM grant IN115720.
\bibliographystyle{abbrv}
\bibliography{Bibliography}

\end{document}